# A processing and analytics system for microscopy data workflows: the Pycroscopy ecosystem of packages

R. Vasudevan*,[1], S. M. Valleti[2], M. Ziatdinov[1,3], G. Duscher[4],**, S. Somnath[5,6]

[1]Center for Nanophase Materials Sciences, Oak Ridge National Laboratory

[2]Bredesen Center for Interdisciplinary Research, University of Tennessee, Knoxville

[3]Computational Sciences and Engineering Division, Oak Ridge National Laboratory

[4]Department of Materials Science and Engineering, University of Tennessee, Knoxville

[5]National Center for Computational Sciences, Oak Ridge National Laboratory

[6]Current affiliation: McKinsey & Company, 280 Congress Street Suite 1100, Boston, MA 02210

**Abstract**

Major advancements in fields as diverse as biology and quantum computing have relied on a multitude of microscopic techniques. All optical, electron and scanning probe microscopy advanced with new detector technologies and integration of spectroscopy, imaging and diffraction. Despite the considerable proliferation of these instruments significant bottlenecks remain in terms of processing, analysis, storage, and retrieval of acquired datasets. Aside from lack of file standards, individual domain-specific analysis packages are often disjoint from the underlying datasets, and thus keeping track of analysis and processing steps remains tedious for the end-user, hampering reproducibility. Here, we introduce the pycroscopy ecosystem of packages, an open-source python-based ecosystem underpinned by a common data model. Our data model, termed the N-dimensional spectral imaging data format, is realized in pycroscopy's *sidpy* package. This package is built on top of dask arrays, thus leveraging dask array attributes, but expanding them to accelerate microscopy relevant analysis and visualization. Several examples of the use of the pycroscopy ecosystem to create workflows for data ingestion and analysis are shown. Adoption of such standardized routines will be critical to usher in the next generation of autonomous instruments where processing, computation and meta-data storage will be critical to overall experimental operations.

*Corresponding authors*

*vasudevanrk@ornl.gov

**gduscher@utk.edu

**Introduction**

Structural, functional and chemical imaging with microscopy have been at the forefront of innovations and discoveries in a wide range of fields in the past three decades and have been critical enabling instruments of the nanotechnology revolution.(Bian, et al., 2021; Fernandez-Leiro & Scheres, 2016; Pennycook & Nellist, 2011) Indeed, microscopy has been able to visualize materials and their associated dynamics across multiple length and time scales. This includes atomic-scale studies of everything from spike protein associated with the SARS-CoV-2 virus(Zhu, et al., 2021) to individual oxygen adatoms hopping on managanite surfaces.(Bryant, et al., 2011; Vasudevan, et al., 2015b) In many microscopes, the ability to perform local spectroscopy is a key advantage, and it can enable e.g., the measurement of local chemistry, electronic states, or functional properties at that site.(Balke, et al., 2015; Newbury & Ritchie, 2013) Those data can then be correlated with microstructural features of the sample obtained from images. Such a process is critical to provide insights into structure-property relationships within the measured sample.(Gázquez, et al., 2017)

Despite the proliferation of different types of microscopes and their considerable use across many scientific domains, significant bottlenecks are now presented by the acquisition, storage, processing and analysis requirements of many modern microscopes.(Kalinin, et al., 2016; Spurgeon, et al., 2021) Even for comparatively 'small-data' microscope platforms, such as atomic force microscopies (AFM), recent advances in terms of newly developed multi-modal spectroscopies, and the ability to acquire streams of uncompressed data directly from the sensor, have greatly increased possible data sizes and data rates, with individual datasets >100GB now eminently possible.(Collins, et al., 2016; Somnath, et al., 2016) For electron microscopy, the problem is compounded, with proliferation of fast detectors and modalities such as 4-dimensional scanning transmission electron microscopy (4D-STEM), which are capable of multiple TB of data capture in a single session.(Ophus, 2019; Ophus, et al., 2014) The sudden jump in data rates and data sizes meant that traditional workflows for analysis, which relied on desktop computers and often vendor-written analysis packages, are no longer viable. Moreover, much of modern microscopy has advanced with applications of deep learning(Ziatdinov, et al., 2017) and other machine learning methods(Schmidt, et al., 2019), assisting everything from data cleaning(Lee, et al., 2020), to improved functional fits(Borodinov, et al., 2019), image segmentation(Bermúdez-Chacón, et al., 2018), and recently, to autonomous microscopy operations.(Kalinin, et al., 2021b;

Ziatdinov, et al., 2022) Therefore, datasets should be able to be used in distributed systems or supercomputer centers. As such, with the recent advancement in automated platforms, standardizing the data model and processing pipelines is an imperative for maximizing the utility of microscopy, and for reducing the significant bottlenecks in the analysis, storage, processing and retrieval that are currently within traditional experimental paradigms.

To date, there have been several open-source efforts to assist with this problem, either by providing tools for the visualization and processing of spectral and imaging data, or storage and file handling, or both. They are too numerous to list here, and vary significantly by discipline, but some examples known to the authors include HyperSpy(de la Peña, et al., 2017), which focuses on hyperspectral data traditionally acquired on electron microscopy platforms, Hystorian(Musy, et al., 2021), a package used for storing imaging data from atomic force microscopy, and LiberTEM(Clausen, et al., 2020), a package to perform distributed computations common in the STEM community.

In contrast to the aforementioned efforts, which are all individual packages, here we introduce the pycroscopy ecosystem of packages, to handle in modular fashion, all parts from acquisition to processing. Also, the package includes conversion from vendor-specified formats to an open data model, visualization, processing, and file writing. Our core philosophy has six tenets: (1) **_Modularity_**, meaning that we provide a set of interoperable and extensible, modular packages, providing a low barrier of entry and high flexibility; (2) **_Consistency_**, in that the core packages can all accept and return *sidpy.Dataset* objects (discussed later); (3) **_Accessible_**, for both users and developers with one-line methods for visualization, creation and saving of data; (4) **_Scalable_**, with data written to the hierarchical data format(Folk, et al., 2011) (HDF5), and dataset objects built on top of Dask(Rocklin, 2015) arrays, thus capable of leveraging chunking and not being memory-limited; (5) **_Standardized_**, with data models that are open and capable of representing data regardless of origin, dimensionality and size, and analysis is not tied to specific data types; and (6) **_Traceable_**, such that raw, intermediate and final data can be written to the same data file and linked to capture provenance.

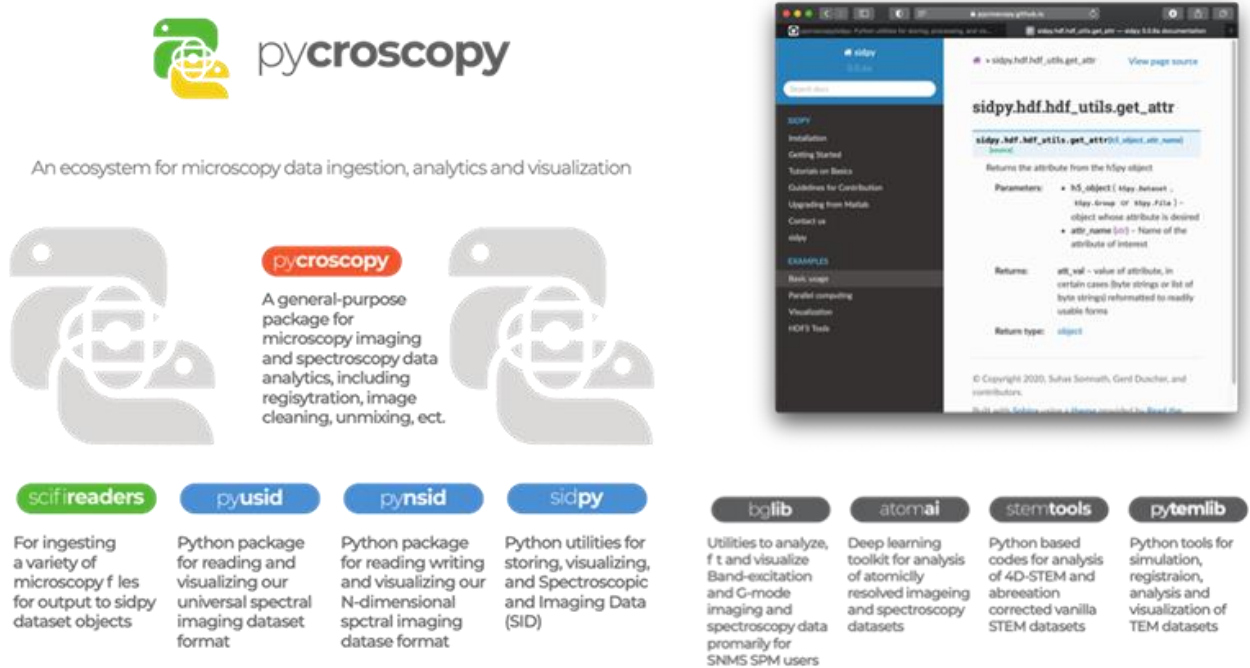

**Figure 1: The pycroscopy ecosystem.** Green/Blue packages (SciFiReaders, Sidpy, pyNSID) are involved with data input, output and visualization. Domain-specific tools like AtomAI and BGLib (grey) are utilized for more specialized analysis or domain tasks.

To realize this philosophy, we crafted interoperable python packages that fall into one of three distinct areas, namely *(1)* Data and file I/O and associated utilities, *(2)* General imaging and spectral utilities, and *(3)* Domain-specific packages, as shown in **Figure 1**. The data I/O utilities are used for describing the data model, and also includes the SciFiReaders package for ingestion of data from a variety of vendor formats to output to the *sidpy.Dataset* object. A general imaging and spectral analysis package is provided, and this package is also named pycroscopy. Finally, domain-specific packages such as BGlib and pyTEMlib slide into the ecosystem for specific, bespoke analysis routines that are generally only used within their respective domains.

Data Model: Sidpy and NSID

The key element of the ecosystem is the *sidpy.Dataset* object. This is a python object that extends the array object of the Dask package, with specific additions to enable writing of meta-data, handling of dimensions with quantities and units, and extended methods for plotting and intelligence folding (reshaping). The sidpy dataset object represents data through the N-dimensional spectral imaging dataset (NSID) model. This is a generic model that takes any N-dimensional matrix of numerical data, in contrast with the universal spectral imaging dataset (USID) model, which we have previously published. Although the NSID model is not capable of representing as many use cases as the original USID model(Somnath, et al., 2019) (it can only be used when an N-dimensional representation for the dataset exists), we find it provides a good balance between ease of use and generality. The N-dimensional matrix comprises the numerical data for the *sidpy* dataset, and additional information on each of the N-dimensions is provided by the user, both in terms of the data itself for the dimension, the type of dimension (e.g., 'spatial', or 'spectral') as well as labels regarding the units and quantity being measured. These are handled internally as a dictionary housed within the dataset object, under the dataset._axes attribute. Finally, each dataset created can be specified with a data type such as 'SPECTRUM', 'SPECTRAL_IMAGE' or 'IMAGE_STACK' and 'IMAGE_4D'. This datatype attribute assists with knowing which plotting function should be utilized when the dataset's plot() method is called. If the dataset object was created from data coming from an instrument, the meta-data will be stored within the dataset.original_metadata attribute as a dictionary. Subsequently, any additional metadata (e.g., associated with data processing) can be stored within the dataset.metadata attribute. Thus, the dataset object provides the ability to represent generic imaging and spectral data, visualize the data, access the associated meta-data, and understand the parameters for each axis.

The sidpy.Dataset object is summarized in **Figure 2a.** The overall philosophy of the pycroscopy ecosystem is shown in **Figure 2b**. To provide insights into the capabilities of the ecosystem, we proceed to provide several use-cases which employ multiple parts of the ecosystem, highlighting its features and advantages.

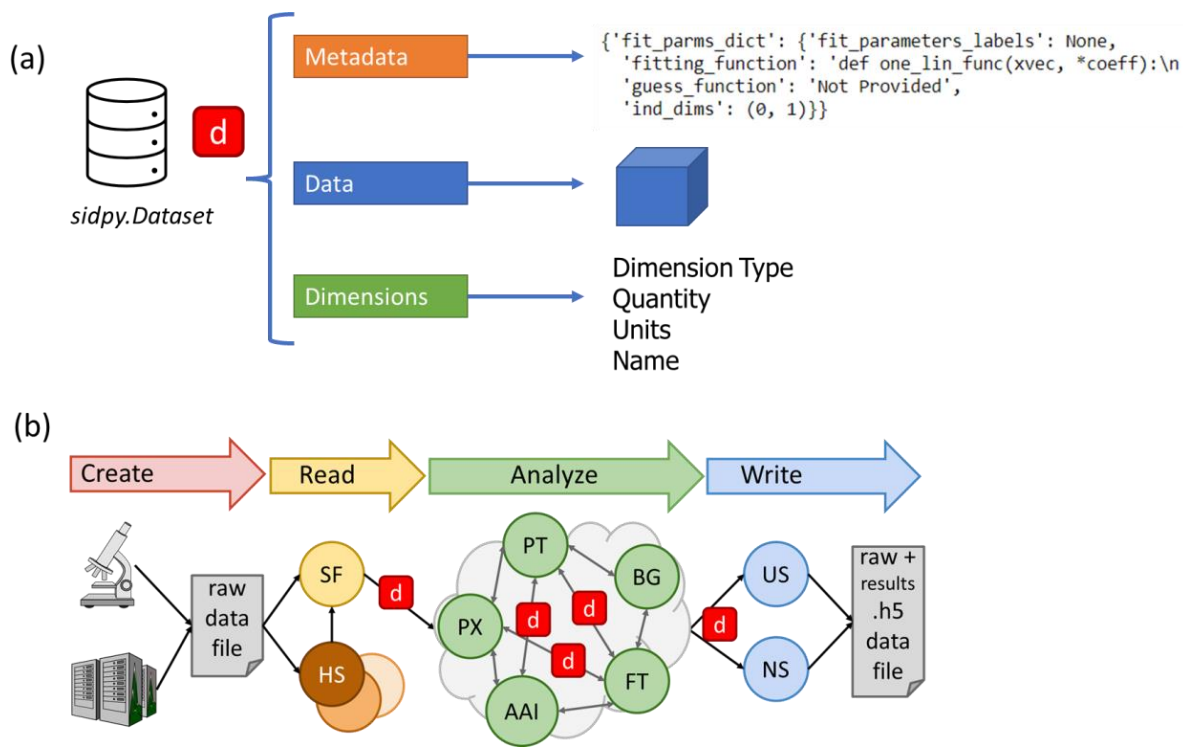

**Figure 2:** Sidpy dataset and the ecosystem. (a) Sidpy.Dataset objects contain data, metadata and dimension information. (b) The overall philosophy of the pycrosopy ecosystem is one in which data is the sidpy.Dataset object is the common 'currency' amongst the packages. Special utilities exist for reading data into these objects, and for saving them to disk (currently we utilize HDF5 files to achieve this).

Workflow example 1: Texture analysis on a STEM image

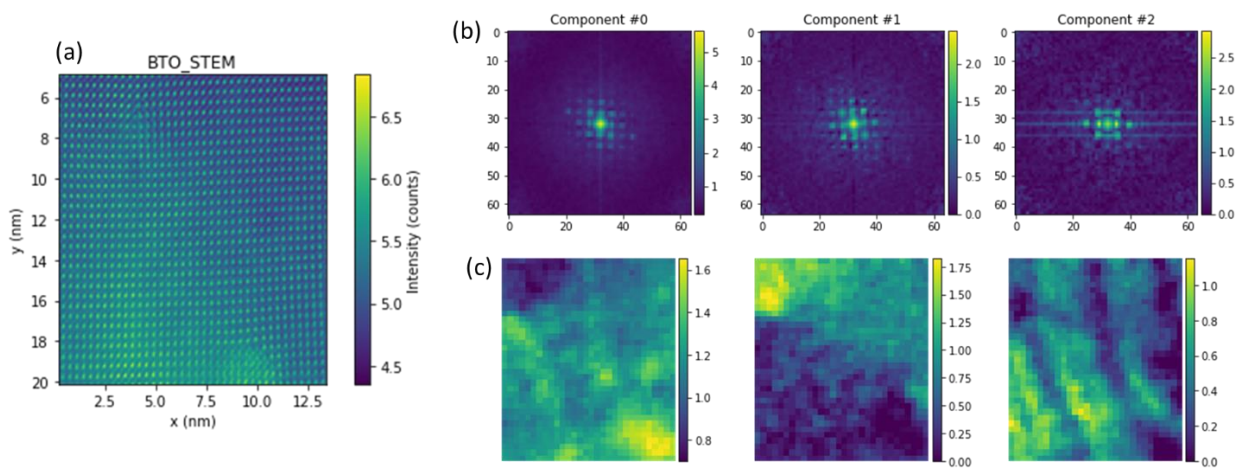

**Figure 3:** Image windowing and matrix factorization. (a) cropped STEM image of BaTiO$_3$. After application of a sliding window Fourier transform, we can perform matrix factorization on the dataset. In this case, non-negative matrix factorization was the method utilized, with the components and abundances shown in (b) and (c), respectively. As reported previously(Kelley, et al., 2022), banded structures appear in the third component, which was suggested to be due to oxygen vacancies.

The first task after acquisition of the imaging and/or spectral data is to convert the data to a usable form. For the pycroscopy ecosystem, this means that we must convert from the vendor-supplied form to sidpy.Dataset objects. We term this process 'reading', and developed the python package 'SciFiReaders' that enables ingestion of data from a wide variety of formats, and reads them into sidpy.Dataset objects. This object can then be used within the rest of the pycroscopy ecosystem and leverage the advantages that the sidpy dataset object provides, that are considerable. As an example, consider the workflow shown in Figure 3. Here, we show previously published data of atomically resolved data $BaTiO_3$ thin film acquired with scanning transmission electron microscopy.(Kelley, et al., 2022) These were originally captured in the dm3 file format. We instantiate the DM3reader, and call the read() method to convert the file into a dataset object. By using the .plot() method in an interactive jupyter notebook, we can immediately plot the data with any axis information available in the file. To analyze only the $BaTiO_3$, and not the substrate, we cropped the data, writing the data back to a sidpy dataset object. The cropped image is shown in Fig. 3(a).

Next, we perform a texture analysis of the cropped region, with a sliding window transform, described in refs(Vasudevan, et al., 2015a). Briefly, patches of the image are created by sliding a window of a set size across the image, such that in the resulting dataset each pixel is associated with the image patch centered on that pixel's location. This is accomplished by calling the generic pycroscopy package, which houses the ImageWindowing() class. The output of this process is a 4-D dataset, but crucially, is still a *sidpy.Dataset* object. Therefore, it can also be immediately visualized with an interactive visualizer simply by calling the dataset's plot() method. This is extremely useful for novice and seasoned users alike, as it accelerates the visualization process considerably. Within the image windowing method, there is the option to choose to perform a local 2D Fourier transform (FFT). In this instance, we used the 2D FFT on each window patch, as we wish to study the changes in symmetry via a matrix factorization method (explained below).

There are several reasons for image windowing. One reason is that it can be used for denoising, by utilizing principal component analysis or some other matrix factorization method, and reconstructing the image patches with a smaller subset of components, before re-stitching the image from the windows, as we have shown previously.(Somnath, et al., 2018) Alternatively, image patches may be used for training of neural networks, to correlate local structure to local functionality (i.e., as measured through spectra).(Kalinin, et al., 2021a; Roccapriore, et al., 2021) Thirdly, these image stacks can also be analyzed through a combination of 2D FFT of each image patch, followed by techniques such as independent component analysis or non-negative matrix factorization (NMF). As we have shown previously, this enables determination of structural phases within atomically resolved images, and maps their spatial positions as well as their strengths.(Vasudevan, et al., 2016) To facilitate this, the pycroscopy package has a MatrixFactorization class, which takes as input the sidpy dataset object and performs the chosen decomposition, before providing the result (the component and abundance maps) as separate (sidpy) datasets. Critically, the factorization works only in 2-dimensions (spatial x spectral), thus the dataset needs to be reshaped according to this expected model. Internally, sidpy contains an intelligent 'fold' method that locates the spatial and spectral datasets and folds any N-dimensional dataset that contains 2 spatial and (N-2) spectral dimensions, into the format required for matrix factorization. The non-negative matrix factorization method yields components and abundance maps, and these can be visualized in the notebook, and are shown in Fig. 3(b,c). In this instance, component 3 appears to show banded features where the tetragonality differs, likely due to the presence of oxygen vacancies (note that this data and analysis was used in a published report by a subset of the authors recently(Kelley, et al., 2022)).

Workflow example 2: Parallel spectral fitting of current-voltage curves

As a second example, we show a routine situation of functional fitting of spectra. In many spectroscopic datasets, a general workflow is to capture the data, visualize it, perform some spectral fitting, and then visualize the fit parameters. Here, we focus on current-voltage curves taken on a sample of $BiFeO_3$, as reported previously in ref(Vasudevan, et al., 2017). The data is read from a HDF5 file into a sidpy dataset, and the plot() method enables point and click visualization of the spectra. This is shown in Fig. 4(a).We proceed to perform spectral fitting by calling on sidpy's SidFitter class. This class takes as input the sidpy dataset, the function to be fit, and a few other arguments. It is also possible to provide a 'guess' function, i.e. a function that will be used to generate the priors for the actual functional fit. Calling the fit() method leverages the parallelism of the underlying Dask framework to split the fitting tasks amongst the stipulated number of workers. Since this is handled at the backend, the user does not need to know anything about parallelism for this considerable speedup. We note here also that the SidFitter class offers the capability to use a 'prior generation' strategy that we have reported previously.(Creange, et al., 2021) Briefly, it involves performing k-means clustering on the spectral data, and then performing functional fits on the cluster means. The parameters of the fit are then taken as priors for the individual members of each cluster. This results in both a speedup (because the optimization is likely to require less iterations), as well as superior results, due to avoidance of situations where poor priors lead to local minima where the optimizer becomes trapped. The results can then be plotted using the visualize_fit_results() method of the fitter class. This is shown in Fig. 4(b).

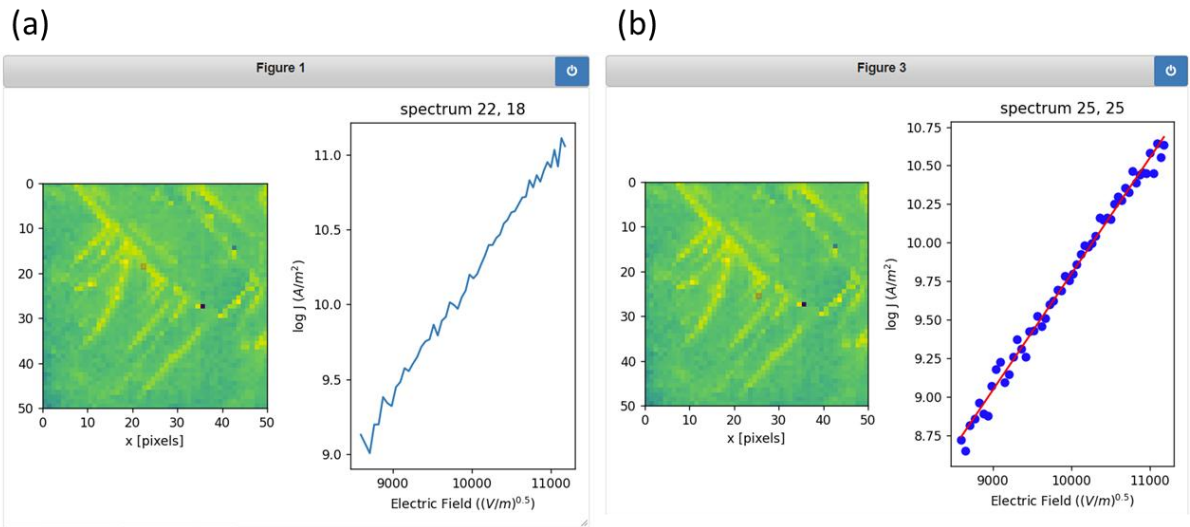

**Figure 4:** Curve fitting example. I-V data from a BiFeO$_3$ thin film sample (data from ref(Vasudevan, et al., 2017)) is shown and can be plotted and viewed interactively in the Jupyter notebook using the .plot() command. After defining the function to be fit, calling the SidFitter class enables parallel fitting, and returns another sidpy dataset object with the fit results. These can be visualized overlaid with the original data, as in (b).

Workflow 3 example: Image registration

A very common task with imaging data is the need to perform image registration. Here, we performed image registration using the pyTEMlib package. First, we import an image stack and display the stack of images. The built-in visualizer enables using the mouse wheel to scroll between the different images in the stack, in this case the images are a scanning transmission electron microscopy image stack of SrTiO$_3$. As there is a non-zero drift, images captured show some minor shifts from one image to the next. We can then perform either rigid or non-rigid registration to align all the images in the stack. Shown in Fig. 5(a) is the HAADF STEM image of SrTiO$_3$. PyTEMLib calls on the simpleITK package to implement Diffeomorphic Demon Non-Rigid Registration. The output is once again a sidpy dataset, and the results show that after registration, the difference between one image and the next is imperceptible. A fit of the drift over time is shown in Fig. 5(b). PyTEMLib can further be used to find the atomic positions with a simple blob finder after application of the Lucy-Richardson deconvolution method. Subsequent fitting enables refinement of the positions to sub-pixel accuracy, and file tools ensure that the data is written back to the h5 file. Note that it is also possible to perform local crystallography with the refined atomic positions, through pycroscopy's Atoms module, but this is not explored here.

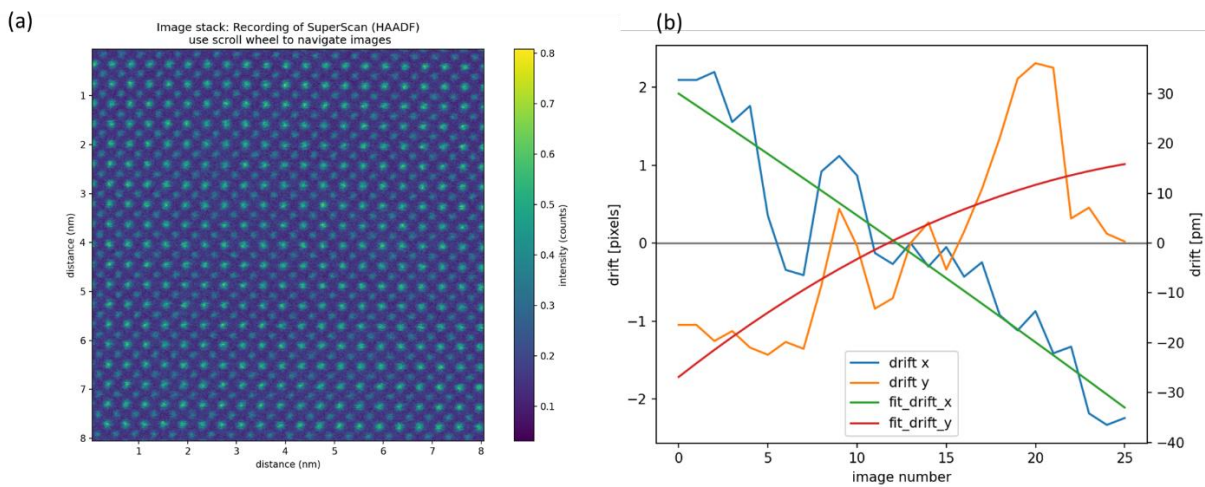

**Figure 5:** Image registration in pyTEMLib. (a) Image stack of SrTiO$_3$ STEM images. In this stack, there are 25 images. (b) After image registration, the drift in the x and y directions and their respective fits are shown in (b).

Workflow 4 example: EELS Spectral analysis

Finally, we look to the example of spectral analysis for core-loss spectra, a feature provided by PyTEMLib. Electron energy loss spectroscopy (EELS) can enable quantitative chemical mapping of samples within the scanning transmission electron microscope. PyTEMLib enables analysis of EELS spectra by providing a graphical user interface built within the Jupyter notebook to accomplish this task. First, spectra can be visualized as shown in Fig. 6(a), and then cropped to the relevant energy scale for subsequent processing. The user can then select the elements present in the sample, which are then used for the fitting process, as well as input other relevant parameters, which is illustrated in Fig. 6(b). The 'fit composition' button enables fitting of the spectral peaks, and performs the calculation of the composition from this fit. All of this is possible within pyTEMLib through the pop-up GUIs written within PyTEMLib, and the data can then be saved back into the hdf5 file to ensure reproducibility. The fit window and result are shown in Fig. 6(c) for illustration, but the user is encouraged to explore this in the interactive Jupyter notebook.

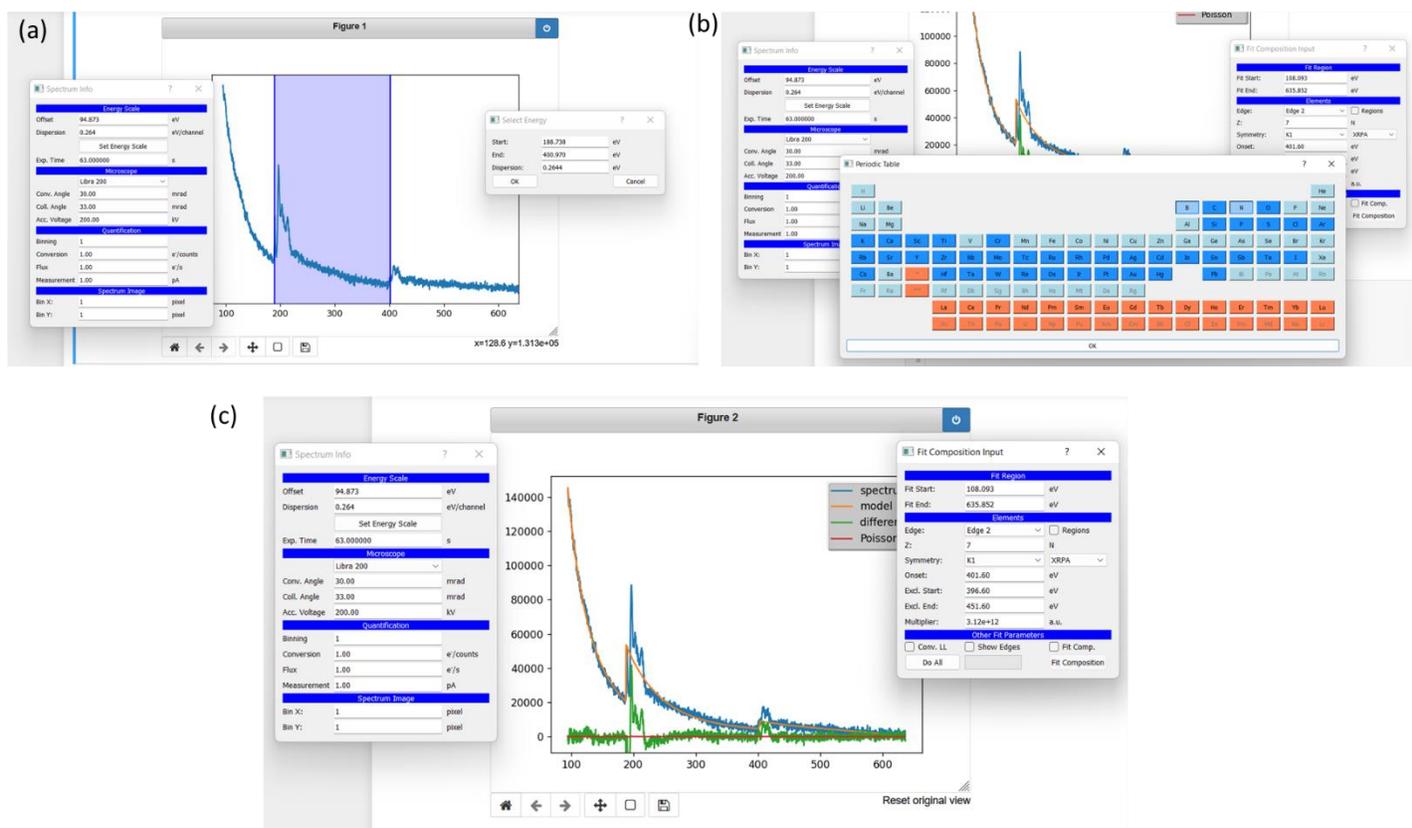

**Figure 6:** GUIs in pyTEMLib for analysis of STEM-EELS datasets. (a) the region of the spectrum can be selected for further analysis. (b) The user can then select the elements in the sample which will enable peak fitting. The composition can then be determined as shown in the 'fit composition' window, in (c).

Discussion

The aforementioned examples illustrate how the pycroscopy ecosystem of packages can be effectively used to generate reproducible scientific data analysis workflows for microscopy datasets. One of the key advantages of the system is the intrinsic methods written into the sidpy dataset object for parallel fitting, visualization, and metadata storage, and ability to write it to disk quickly. With the rise of automated and autonomous instruments(Hammer, et al., 2021; Kalinin, et al., 2021b) and cloud laboratories(Arnold, 2022) reproducible and standardized workflows are required, which necessitate software and hardware infrastructure to support these efforts. For instance, training of machine learning models for defect detection in microscopy images implies the need to search through large volumes of data for retrieving and creating training datasets. Such a search would be simplified if all the data in a lab or research facility would be standardized with ready access to metadata. Within our own facility, HDF5 files that follow either the NSID or USID paradigm are able to be tagged and stored within databases managed by an in-house(and freely available) software solution termed DataFed.(Stansberry, et al., 2019) Images and datasets acquired at the instrument can be pushed to centralized repositories and processing pipelines can be immediately triggered in the cloud, to ensure that the data is read and converted into the standard sidpy dataset, all meta-data is extracted, a DataFed record is created, and that certain processing steps are also run, depending on the file and experiment type.

With the rise of more automated experimental platforms in both synthesis and characterization, we envision the workflows to become more complex and involve multiple instruments. As such, future efforts within the pycroscopy ecosystem should focus not only on adding additional machine learning functionality and improved integration of the different packages currently available, but also to improve schema for recording the provenance of operations. Overall, we find the current system provides a strong degree of flexibility as opposed to a more rigid one, and thus reduces the barriers for those wanting to adopt the platform for their data analysis needs.

**Acknowledgements**

This work was supported by the Center for Nanophase Materials Sciences, which is a US Department of Energy, Office of Science User Facility at Oak Ridge National Laboratory.

**Conflict of Interests Statement**

The authors declare no conflicts of interest

**Data and Code**

Jupyter notebooks for pycroscopy to reproduce the results here are available in https://github.com/pycroscopy/pycroscopy. Those pertaining to pyTEMlib are available in https://github.com/pycroscopy/pyTEMlib